\documentclass[reprint,amsmath,amssymb]{revtex4-2}
\usepackage{graphicx}
\usepackage{amsthm}
\usepackage[shortlabels]{enumitem}
\usepackage{braket}
\usepackage{appendix}
\usepackage{mathtools}
\usepackage{tikz}
\usepackage{yquant}
\usepackage{yquant}

\theoremstyle{plain}
\newtheorem{thm}{Theorem}

\newtheorem*{remark}{Remark}

\begin{document}

\title{The Problem with Grover-Rudolph State Preparation \\ for Quantum Monte-Carlo Integration}
\author{Steven Herbert}
\email{Steven.Herbert@cambridgequantum.com}
\affiliation{Cambridge  Quantum  Computing  Ltd,  9a  Bridge  Street,  Cambridge,  CB2  1UB,  UK }
\affiliation{Department of Computer Science and Technology, University of Cambridge, UK}

 \begin{abstract}
    \noindent 
    We prove that there is no quantum speed-up when using quantum Monte-Carlo integration to estimate the mean (and other moments) of analytically-defined log-concave probability distributions prepared as quantum states using the Grover-Rudolph method.
\end{abstract}

\maketitle

\section{Introduction}
\label{intro}

A frequent, and much relied-upon claim in quantum finance and quantum data-science literature is that there is an automatic quadratic quantum advantage when performing Monte-Carlo integration to approximate the expectation of some probability distribution, if the distribution is efficiently integrable \cite{Rebentrost2018, Woerner2019, Stamatopoulos_2020, QCfinance, egger2019credit, chakrabarti2020threshold, rebentrost2018quantum, kaneko2020quantum, financeppr}. In this paper we show that this is not true in general. The crucial factor is that the quadratic speed-up arises from comparing classical complexity to quantum \textit{query} complexity, and if we take into account the additional operations needed to build increasingly precise oracles for the quantum Monte-Carlo (i.e., as the overall accuracy required increases), then we find that there is no quantum advantage.\\
\indent To see this, we can start by looking at the essential problem that Monte-Carlo integration solves. If we have some probability density function $p(x)$, for which we want to calculate the expectation then we must calculate the integral:
\begin{equation}
\label{eqn10}
    \mathbb{E}(x) = \int_x x p(x) \, \mathrm{d}x
\end{equation}
However, should this integral not be analytically calculable, then we may resort to Monte-Carlo integration. This entails sampling from $p(x)$ a number (say $N_s$) times, and averaging the samples to approximate the expectation:
\begin{equation}
    \label{eqn20}
    \mathbb{E}(x) \approx \frac{1}{N_s} \sum_{j=1}^{N_s} x_j
\end{equation}
where $x_j$ are samples from $p(x)$. In this case, the approximation error is well-defined: the root mean squared error (RMSE) is proportional to $1/\sqrt{N_s}$ \cite{MCbook}.\\
\indent If, however, we are in possession of some quantum circuit that prepares the $n$-qubit state:
\begin{equation}
\label{eqn30}
    \ket{\psi_n} = \sum_{i \in \{0,1 \}^n}\sqrt{\mathrm{p}_i}\ket{i}
\end{equation}
where $\{ \mathrm{p}_i \}$ is a $2^n$-point discretisation of $p(x)$, then we can use the \textit{quantum} Monte-Carlo to achieve RMSE that decays as $\Theta(1/N_q)$, where $N_q$ is the number of times that the circuit preparing $\ket{\psi_n}$ is queried as part of an oracle circuit \cite{brassard2000quantum, Suzuki_2020, grinko2019iterative, MontanaroMC, Aaronson_2020, nakaji2020faster}. That is, a quadratic speed-up in query complexity, where a classical query involves one sample from $p(x)$, and a quantum query involves one oracle call to the state preparation circuit.\\
\indent This quadratic quantum advantage can indeed be realised if we are in possession of a circuit that \textit{exactly} encodes $\ket{\psi_n}$. However, in general, the quantum circuit will only \textit{approximately} prepare $\ket{\psi_n}$, and furthermore in order to more accurately prepare $\ket{\psi_n}$, the process of state preparation will itself generally be more computationally expensive.\\
\indent In this paper we address the case where the Grover-Rudolph method \cite{grover2002creating} is used to prepare $\ket{\psi_n}$ encoding some log-concave distribution \cite{logconcave}. Notably, many very commonly-used distributions such as the exponential and normal families are log-concave. The Grover-Rudolph method takes $\Theta(n)$ steps to prepare a discrete
distribution with $N=2^n$ points, and thus is efficient in this sense --
with the important caveat that $p(x)$ need to be efficiently
integrable. The authors claim that the efficiency holds, even when the
integration is performed numerically, so long as an \textit{efficient}
approximate integration algorithm exists. This claim is, of course,
almost a tautology, but there is an important subtlety here:
In their paper, Grover and Rudolph implicitly take ``efficiently'' to
mean ``with only a polynomial time overhead in the problem size'',
which would obviously preserve any \textit{exponential} speed-ups in
algorithms that take $\ket{\psi_n}$ as an input. However, in the case
where quantum Monte-Carlo takes $\ket{\psi_n}$ as an input, only a
\textit{quadratic} speed-up is available, and so it is necessary to
demand an accordingly stricter standard for what ``efficiently
integrable'' should actually mean. The central claim of this paper is
that when classical Monte-Carlo integration is used in the state
preparation circuit then this \textit{is not} sufficiently efficient
to preserve the quadratic quantum advantage in quantum Monte-Carlo.
%

\section{Preliminaries}

For simplicity, we assume that the distribution of interest has been discretised over $N=2^n$
points (for some integer $n$) and shifted and scaled such that its domain is $\{ 0 , 1
, \dots, 2^{n}-1 \}$. We let $\mu$ be the mean of $p(x)$ and $\hat{\mu} = \mathbb{E}(x)$ be an estimate of $\mu$. As mentioned in Section~\ref{intro} the natural measure of error for Monte-Carlo integration is the RMSE, defined:
\begin{equation}
\label{eqn40}
    \hat{\epsilon} = \sqrt{\mathbb{E}((\hat{\mu}- \mu)^2)}
\end{equation}

\subsection*{Grover-Rudolph State Preparation}

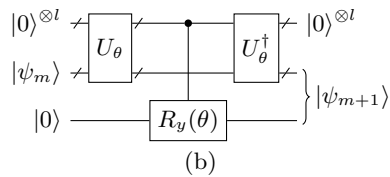
\begin{figure}[!t]
  \centering

\begin{tikzpicture}
  \begin{yquant}
    qubit {$\ket{0}$} q[1];
    box {$R_y(\theta_0)$} q[0];
    output {$\cos \theta_0 \ket{0} + \sin \theta_0 \ket{1}$} q;
  \end{yquant}
\end{tikzpicture}\\
    (a)\\
  \vspace{0.5cm}
  \begin{tikzpicture}
    \begin{yquant}
    qubit {$\ket{0}^{\otimes l}$} r[1];
    qubit {$\ket{\psi_m}$} q[1]; 
    qubit {$\ket{0}$} s[1];
    slash q[0];
    slash r[0];
    box {$U_\theta$} (q[0],r[0]) ;
    slash q[0];
    slash r[0];
    box {$R_y(\theta)$} s[0] | r[0];
    box {$U^\dagger_\theta$} (q[0],r[0]) ;
    slash q[0];
    slash r[0];
    output {$\ket{0}^{\otimes l}$} r[0];
    output {$\ket{\psi_{m+1}}$} (q[0],s[0]); 
  \end{yquant}
  \end{tikzpicture}

  (b)\\
  
  
  \caption{Quantum circuits for the Grover-Rudolph method: (a) for the first iteration; and (b) for subsequent iterations, where $\ket{0}^{\otimes l} \ket{i} \xrightarrow[]{U_\theta}  \ket{\theta_i} \ket{i}$. Note that $l$ is the number of qubits needed to represent $\theta$ to appropriate accuracy; and also that the controlled $R_y(\theta)$ gate shown is shorthand for a series of $R_y$ gates with fixed rotation angles, each conditionally controlled by the various qubits in the $\theta$ register, such that the overall action is a rotation by $\theta$.}
  \label{f1}
\end{figure}

The Grover-Rudolph method constructs the state $\ket\psi$ iteratively,
at each step adding a qubit and doubling the number of points in the
discrete probability distribution.  Since each point in the discrete
distribution corresponds to an interval, the new distribution is
obtained by splitting each interval in two; the allocation of
probability mass to the left and right halves is computed by
integration.
Moreover, the quantum-nature of the
algorithm allows this to happen for all points in superposition (i.e.,
in a single step). This is achieved by applying the rotation which has
the following effect on the $i^{th}$ computational basis state:
\begin{equation}
\label{eqn110}
     \ket{\theta_i} \ket{i} \ket{0} \to  \ket{\theta_i} \ket{i} (\cos \theta_i \ket{0} + \sin \theta_i \ket{1} ) 
\end{equation}
where:
\begin{equation}
\label{eqn120}
    \theta_i = \arccos \sqrt{f(i)}
\end{equation}
and:
\begin{equation}
\label{eqn130}
    f(i) = \frac{ \int_{x_L^i}^{\frac{x_R^i - x_L^i}{2}} p(x) \, \mathrm{d} x} { \int_{x_L^i}^{x_R^i} p(x) \, \mathrm{d} x}
\end{equation}
where $x_L^i$ and $x_R^i$ are the left- and right-hand boundaries of the $i^{th}$ interval respectively. The very first Grover-Rudolph iteration is an unconditional rotation, whose circuit is shown in Fig.~\ref{f1}(a). The circuit for the subsequent iterations is shown in Fig.~\ref{f1}(b) -- in particular, the iterations of the Grover-Rudolph method are indexed ``$m$'', and Fig.~\ref{f1}(b) shows the circuit for adding the $(m+1)^{th}$ qubit to the state. \\
%
\indent The Grover-Rudolph paper simply states that this procedure readily generalises to higher dimensions, however we give an explicit construction to this effect in Appendix~\ref{app0}. To simplify the following analysis, we consider only a univariate probability distribution, however we note that Monte-Carlo integration is unlikely to be the method of numerical integration of choice in this setting. Therefore in Appendix~\ref{app0} we also briefly discuss why the following main result will trivially extend to the case of using quantum Monte-Carlo integration to estimate the mean of a multivariate distribution, when the Grover-Rudolph method is used to prepare the state encoding that distribution.

%

\section{Main Result}

\noindent As stated in Section~\ref{intro}, the RMSE of a classical Monte-Carlo estimate of the mean is proportional to $\frac{1}{N_s}$, which corresponds to a sample complexity $N_s \in \Theta \left( \frac{1}{\hat{\epsilon}^2} \right)$. We show in Appendix~\ref{app} that, when implemented on a digital classical computer this corresponds to a computational complexity of:
\begin{equation}
    \tilde{\Theta}\left( \frac{1}{\hat{\epsilon}^2} \right)
\end{equation}
That is, the computational complexity is only a negligible poly-logarithmic factor greater than the sample complexity (as indicated by the tilde). We now show that there is no quantum advantage if the Grover-Rudolph method is used to prepare a state for quantum Monte-Carlo.\\


%
%

\begin{thm}
To achieve a RMSE of $\hat{\epsilon}$ using an unbiased quantum Monte-Carlo estimation method requires $\tilde{\Omega}\left(\frac{1}{\hat{\epsilon}^{2}}\right)$ operations when the Grover-Rudolph method is used to prepare some log-concave distribution as a quantum state.
\end{thm}

\begin{remark}
The requirement that the quantum Monte-Carlo estimation is unbiased is included purely for technical reasons in the following proof. In practice, even a biased estimator would yield no advantage, unless that bias were to be correlated with the error of an ostensibly independent classical Monte-Carlo estimation of some aspect of the distribution, in such a way the two errors cancel out. It is, however, worth noting that quantum Monte-Carlo techniques based on phase-estimation are likely to be unbiased in general; and also the NISQ-appropriate algorithm ``amplitude estimation without phase estimation'' claims to be asymptotically unbiased \cite{Suzuki_2020}.
\end{remark}

\begin{proof}
The proof strategy is thus: firstly, we assume an error is present in the preparation of $\ket{\psi_n}$ and show the effect of this on the RMSE; and secondly we demonstrate this such a state-preparation error \textit{is} present when the Grover-Rudolph method is used, and give its scaling.\\
\indent To assess the convergence rate of quantum Monte-Carlo, we make the conservative assumption that error is only incurred in the first Grover-Rudolph iteration, i.e., the circuit in Fig.~\ref{f1}(a) performs a rotation by an erroneous angle $\theta_0$. Given that the first iteration of Grover-Rudolph calculates the amount of probability mass that is distributed over the left-hand half of the distribution, it is convenient in the following analysis to consider the error in the probability mass function directly. Formally, we let:
\begin{equation}
    \label{neweqn10}
    p_l = \sum_{i=0}^{2^{n-1}-1} \mathrm{p}_i,
\end{equation}
and $\epsilon_l$ be the error incurred in the first Grover-Rudolph iteration, such that rather than preparing state (\ref{eqn30}), we instead prepare the erroneous state:
\begin{equation}
\label{neweqn20}
    \ket{\psi'} =  \sum_{i=0}^{2^{n-1}-1} \sqrt{\frac{\mathrm{p}_i(p_l + \epsilon_l)}{p_l}} \ket{i}  + \sum_{i=2^{n-1}}^{2^{n}-1} \sqrt{\frac{\mathrm{p}_i(1-p_l - \epsilon_l)}{1-p_l}} \ket{i}
\end{equation}
We can now evaluate the mean, $\mu'$, of the erroneous distribution that has been encoded in the state, $\ket{\psi'}$:
\begin{align}
    \mu' & = \sum_{i=0}^{2^{n-1}-1} i \frac{\mathrm{p}_i(p_l + \epsilon_l)}{p_l} + \sum_{i=2^{n-1}}^{2^{n}-1} i \frac{\mathrm{p}_i(1-p_l - \epsilon_l)}{1-p_l} \nonumber \\
    & = \sum_{i=0}^{2^{n}-1} i \mathrm{p}_i + \epsilon_l \left( \sum_{i=0}^{2^{n-1}-1} i \frac{\mathrm{p}_i}{p_l} - \sum_{i=2^{n-1}}^{2^{n}-1} i \frac{\mathrm{p}_i}{1-p_l} \right) \nonumber \\
    \label{neweqn30}
    & = \mu - k \epsilon_l
\end{align}
where $k$ is the mean of the left-hand half of the discretised distribution minus the mean of the right-hand half of the discretised distribution. That is, a non-zero constant that depends only on the distribution being encoded and not the sampling process. We can see that the error, $\epsilon_l$, in the first iteration of the Grover-Rudolph state preparation method is inherited as a proportional term in the error of the distribution mean. It follows that when performing the quantum Monte-Carlo, we are actually sampling from an erroneous distribution whose mean errs from the mean of the actual distribution of interest by a factor proportional to $\epsilon_l$. Thus, we can consider the overall mean squared error, which is equal to:
\begin{align}
    \hat{\epsilon}^2  = & \mathbb{E}\left( \left( \mu - \hat{\mu}(N_q) \right)^2 \right) \nonumber \\
     = & \mathbb{E}\left( \left( \mu' + k\epsilon_l - \hat{\mu}(N_q) \right)^2 \right) \nonumber \\
     = & \mathbb{E}\left( \left( \mu' - \hat{\mu}(N_q) \right)^2 \right) + k^2 \mathbb{E}(\epsilon_l^2) \nonumber \\
     & \,\,\,\, + \mathbb{E}_{\epsilon_l} \left( k \epsilon_l \, \mathbb{E}_{\hat{\mu}| \epsilon_l} \left( \mu' - \hat{\mu}(N_q) \right) \right) \nonumber \\
    \label{eqn250}
     = & \mathbb{E}\left( \left( \mu' - \hat{\mu}(N_q) \right)^2 \right) + k^2 \mathbb{E}(\epsilon_l^2)
\end{align}
where $\hat{\mu}(N_q)$ is the estimate of the mean obtained by performing quantum Monte-Carlo such that the state preparation circuit is queried $N_q$ times. This derivation uses the fact that $\epsilon_l$ is itself a random variable, and that the quantum estimate is unbiased (i.e., by the previous assumption), so $\mathbb{E} \left( \mu' - \hat{\mu}(N_q) \right) = 0$. Thus the effect of (\ref{eqn250}) is to separate the error pertaining to the construction of the state-preparation circuit, and the error pertaining to the inaccuracy of the quantum Monte-Carlo itself. We know that the first term in the sum in (\ref{eqn250}) can be suppressed as $\Theta(1/N_q^2)$. Thus we turn our attention to the second term, which relates to the state preparation.\\
\indent Grover and Rudolph specify that log-concave probability distributions should be numerically integrated using the technique described by Applegate and Kannan \cite{logconcave} which concerns the general case of numerically evaluating the integral over some region of the support of the probability mass. This technique essentially involves dividing the region into halves, and performing Monte-Carlo integration to evaluate the ratio of probability mass in each half -- and continuing recursively (choosing the half with greatest probability mass to further divide each time), until the region is sufficiently small that the probability mass is essentially constant therein. However, we can easily see that this approach can be tweaked slightly in the case in which the integral of the entire area sums to one -- as is the case for the \textit{first} iteration of Grover-Rudolph. In this case, the simpler method of sampling $N'_s$ times and then counting up the proportion that fall in the region of interest can be used -- which is manifestly less computationally complex for given accuracy. We have that the proportion of random samples that will fall in the left-hand half of the distribution is expected to be $p_l$. Letting $X \sim p(x)$ be random samples from the distribution of interest, and $\mathcal{L}$ be the region covered by the left-hand half of the distribution, we can see that the estimate, $\hat{p}_l$ of $p_l$ corresponds to a binomial distribution:
\begin{equation}
    \label{eqn155}
    \hat{p}_l \sim \mathcal{B}(N'_s, p_l)
\end{equation}
We can thus bound the mean squared error, $\mathbb{E}({\epsilon}_{l}^2)$, of this Monte-Carlo estimate of $p_l$:
\begin{align}
    \mathbb{E}({\epsilon}_{l}^2) & = \mathbb{E}\left(\left(p_l - \hat{p}_l\right)^2\right) \nonumber \\
     & = \mathbb{E}\left(\left(p_l - \frac{1}{N'_s}\sum_{X \in \mathcal{L}} 1 \right)^2\right) \nonumber \\
    & = \frac{1}{(N'_s)^2}\mathbb{E}\left(\left(N'_s p_l - \sum_{X \in \mathcal{L}} 1 \right)^2\right) \nonumber \\
    & = \frac{1}{(N'_s)^2}N_s p_l(1- p_l )  \nonumber \\
    \label{eqn160}
     & = \frac{p_l (1- p_l)}{N'_s}
\end{align}
So we can see that, whilst the error in the quantum Monte-Carlo is suppressed as $\Theta(1/N_q^2)$, the error in the classical Monte-Carlo required to prepare the state used in the quantum Monte-Carlo is only suppressed as $\Theta(1/N'_s)$. Returning to (\ref{eqn250}), we can see that the classical Monte-Carlo sampling required to prepare the state sufficiently accurately is therefore the bottleneck, thus giving us an overall computational complexity of:
\begin{equation}
\label{eqn260}
        \tilde{\Omega} \left( \frac{1}{\hat{\epsilon}^2} \right)
\end{equation}
where $\Omega$ is used because this is a lower-bound: we have considered only the error in the first Grover-Rudolph iteration; and furthermore it is $\tilde{\Omega}$, because the classical complexity of the Monte-Carlo sampling will be subject to the same poly-logarithmic overhead as identified in Appendix~\ref{app}.
\end{proof}

\section{Discussion}

At its core, the result we show in this paper is the simple observation that, if classical sampling is required to prepare a quantum state which encodes a probability distribution, then the best estimate of the mean that we can obtain is the average of the samples themselves: putting these samples into and out-of a quantum computer cannot, in and of itself, beget more accuracy. In particular, we have shown that there is no quantum advantage when the Grover-Rudolph method is used to prepare the state for quantum Monte-Carlo.\\ 
\indent For simplicity and definiteness, the analysis in this paper has considered a very restricted error model, which means that the computational overhead required to build a correspondingly more accurate state-preparation circuit as the desired estimation accuracy grows falls solely into a single classical pre-processing step to calculate $\theta_0$ and so the quantum complexity does not increase. In reality, however, the circuit depth of the state-preparation circuit \textit{will} grow with the desired accuracy. This is because the Grover-Rudolph method specifies that the Monte-Carlo integration required in the state preparation should be performed using a register of qubits initialised in a random computational basis state, meaning that the numerical integration is to be achieved using quantum rather than classical operations (in fact this is a necessity in all but the first Grover-Rudolph iteration). It follows that the depth of the circuit $U_\theta$ in Fig.~\ref{f1}(b) will grow with the desired accuracy.\\
%
%
\indent This conclusion adds theoretical basis to the emerging picture that Grover-Rudolph state preparation is too costly in practice \cite[Tab. 1]{chakrabarti2020threshold}. This does not, however, condemn the whole idea of using quantum Monte-Carlo to achieve quantum speed-ups in data-science. For example, even though the Grover-Rudolph paper is very heavily cited, there are alternative ways to prepare the state (eg. Refs.~\cite{vazquez2020efficient, chakrabarti2020threshold}), which may yield a quantum speed-up. Moreover, in cases in which we do not simply want to evaluate the expectation of the probability distribution itself, but rather to first apply function to the random variables before the expectation is to be taken, then there could be a quantum speed-up overall if this function that is applied to the random variables is sufficiently complex. Finally, it is worth noting that, Grover-Rudolph also applies to situations in which the function \textit{can} be analytically integrated, which may occur even when the expectation \textit{cannot} be (that is, functions for which $\int p(x) \mathrm{d}x$ between appropriate limits is analytically calculable, but where $\int x p(x) \mathrm{d}x$ isn't): in this case, the absence of classical Monte-Carlo in the state-preparation circuit means that the quadratic quantum advantage will be upheld, even when considering computational rather than query complexity.\\

\section*{Acknowledgement}

\noindent The author thanks Ross Duncan, Cristina Cirstoiu and Luciana Henaut for their comments and suggestions.



\appendix

\section{Grover-Rudolph for Multivariate probability distributions}
\label{app0}
Suppose we have a $d$-dimensional multivariate probability distribution, $p(x^{(1)}, x^{(2)}, \dots , x^{(d)})$. Assuming each of these dimensions has $2^n$ equally spaced points of probability mass, our aim is to use the Grover-Rudolph method a number of times to prepare a state of the form:
\begin{equation}
\label{app0eqn0}
\ket{\psi} = \sum_{x^{(1)}  \dots \\ x^{(d)} } \sqrt{p(x^{(1)} , \dots , x^{(d)})} \ket{x^{(1)}  \dots x^{(d)}}  
\end{equation}
which thus encodes the multivariate probability distribution by using $d$ registers, each of $n$ qubits, to represent the points of probability mass. The Grover-Rudolph method dictates that $p$ should be efficiently integrable, which in the multivariate case we may take to mean that we can efficiently compute:
\begin{equation}
\label{app0eqn10}
\int_{x^{(1)}_l}^{x^{(1)}_u}  \dots \int_{x^{(d)}_l}^{x^{(d)}_u} p(x^{(1)}, \dots , x^{(d)})  \, \mathrm{d}x^{(1)}  \, \dots \, \mathrm{d}x^{(d)} 
\end{equation}
\begin{figure}[!t]
  \centering
  \begin{tikzpicture}
    \begin{yquant}
    qubit {$\ket{0}^{\otimes n}$} r[1];
    qubit {$\ket{0}^{\otimes n}$} a[1];
    qubit {$\ket{0}^{\otimes n}$} b[1];
    slash r[0];
    slash a[0];
    slash b[0];
    box {GR(1)} (r[0]) ;
    box {GR(2)} a[0] | r[0];
    box {GR(3)} b[0] | a[0], r[0];
    slash r[0];
    slash a[0];
    slash b[0];
  \end{yquant}
  \end{tikzpicture}
  \caption{Grover-Rudolph state preparation for tri-variate distribution. GR(1) denotes the preparation of the marginal distribution of the first dimension of the multivariate distribution (as per normal univariate case of Grover-Rudolph); GR(2) denotes the use of Grover-Rudolph to prepare the conditional distribution of the second dimension given the first (with the third dimension marginalised out) -- the conditioning in the integral achieved by the quantum control; and GR(3) denotes the use of Grover-Rudolph to prepare the conditional distribution of the third dimension given the first and second.}
  \label{f2}
\end{figure}
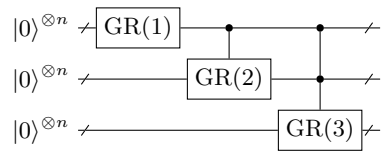
for any limits of integration. In particular, we can compute the integral over the entire support of each dimension, except for one dimension where we set the limits we want -- thus achieving integration of the marginal distribution of the dimension in question. For example, if we wish to integrate the marginal distribution of the first dimension, we get:
\begin{align}
& \int_{x^{(1)}_l}^{x^{(1)}_u}   p(x^{(1)})   \, \mathrm{d}x^{(1)} \nonumber \\
\label{app0eqn20}
& =  \int_{x^{(1)}_l}^{x^{(1)}_u}  \int_{\infty}^{\infty}   \dots \int_{-\infty}^{\infty} p(x^{(1)}, \dots , x^{(d)})  \, \mathrm{d}x^{(1)} \mathrm{d}x^{(2)} \, \dots \, \mathrm{d}x^{(d)} 
\end{align}
where $p(x^{(1)})$ is the marginal distribution of $x^{(1)}$. In order to use the Grover-Rudolph method to prepare a quantum state encoding $p$, we further require that efficient integration of some conditional distributions is possible. In particular, for any $d'$ such that $1 \leq d' <d$ we must be able to efficiently compute the integral:
\begin{align}
\int_{x^{(d'+1)}_l}^{x^{(d'+1)}_u}  \dots \int_{x^{(d)}_l}^{x^{(d)}_u} p(x^{(d'+1)}, \dots ,  x^{(d)} | & x^{(1)}, \dots , x^{(d')})  \nonumber \\
\label{app0eqn30}
& \mathrm{d}x^{(d'+1)}  \, \dots \, \mathrm{d}x^{(d)} 
\end{align}
once again, including the cases where some of the integrals are over the entire domain. We can thus prepare the joint distribution by noticing the decomposition:
\begin{align}
p(x^{(1)}, x^{(2)},  \dots , x^{(d)}) = p(x^{(1)}) & p(x^{(2)} |x^{(1)}) \nonumber \\
&  \dots p'(x^{(d)} | x^{(1)}, \dots , x^{(2)})
\end{align}
and then using Grover-Rudolph a total of $d$ times to prepare first the marginal distribution of $x^{(1)}$, then the conditional distribution of $x^{(2)}$ given $x^{(1)}$ (marginalising out all other variables), through to the conditional distribution of $x^{(d)}$ given $x^{(d-1)} \dots x^{(1)}$. These $d$ uses of the Grover-Rudolph method can be performed coherently in superposition, in exactly the same way that the Grover-Rudolph algorithm itself works. Figure~\ref{f2} gives the circuit diagram to achieve this for the trivariate case.\\
\indent As preparing a $d$-dimensional multivariate probability distribution simply amounts to $d$ uses of the Grover-Rudolph method, we can see that our main result presented for the case of univariate probability distributions will extend to any multivariate probability distribution where numerical integration is used.

\section{Complexity of Classical Monte-Carlo}
\label{app}

In this section we show that, when classical Monte-Carlo is performed on a digital computer, the number of bits required is only poly-logarithmic in the reciprocal of the desired RMSE, and furthermore that all operations have complexity that is polynomial in the \textit{number of bits}. Therefore, performing the classical Monte-Carlo on a digitial computer only introduces a poly-logarithmic overhead in complexity compared to the sample complexity.\\
\indent We start with the definition of the RMSE:
\begin{equation}
\label{eqn170}
\hat{\epsilon}^2 = \mathbb{E}\left( \left( \mu - \frac{1}{N_s} \sum_{j=1}^{N_s} X_j  \right)^2 \right)
\end{equation}
However, when performed on a digital computer, $X_j$ \textit{will not be} samples from the distribution of interest, but rather an approximate version thereof. This approximation arises because of two inaccuracies introduced by performing the sampling on a digitial computer. Firstly, the samples will not be drawn from the (continuous) distribution itself, but rather from some discretised version.


Let the support of $p(x)$ be between $x_l \leq x < x_u$, where the distribution is shifted such that $x_l \geq 0$, and also let $|\frac{\mathrm{d} p(x)}{\mathrm{d} x}| \leq \beta$ for some constant $\beta$ in the region $x_l \leq x < x_u$. Treating $p(x)$ as having only finite support is justified as it can be thought of as the result of a pre-truncating step that is necessary in both the classical and quantum cases.\\
\indent Next, we must define a \textit{discretised} version of $p(x)$ with $N_d$ equally spaced intervals between $x_l$ and $x_u$, for simplicity we let $N_d$ be a power of 2 and also let $\Delta x = x_u-x_l$. 
\begin{equation}
\label{eqn50}
    X = \left\{ x | x = x_l + i \frac{\Delta x}{N_d} \text{ for } i=0 \dots N_d-1 \right\}
\end{equation}
\indent From this, we can define the mean of the discretised distribution:
\begin{equation}
\label{eqn70}
    \mu_d = \frac{1}{N_d} \sum_{i=0}^{N_d - 1} X_i p(X_i)
\end{equation}
This enables us to establish the discretisation error incurred when estimating the error on a digital classical or quantum computer: 
\begin{equation}
\label{eqn60}
\mu_d - \beta/2 \left(\frac{\Delta x}{N_d}\right)^2  \leq \mu \leq  \mu_d + \beta/2 \left(\frac{\Delta x}{N_d}\right)^2
\end{equation}
This is because the definition of $X$ is such that a probability mass function is formed by taking the left-Riemann sum of $p(x)$. So it follows that the maximum discrepancy for each region in the sum is upper-bounded by the area of a right-angle triangle of base $\frac{\Delta x}{N}$ and height $\beta \frac{\Delta x}{N}$. Thus, in the worst case, the probability of each region errs by this maximum amount, and so the mean itself errs by the same. From this we can upper-bound the discretisation error, $\epsilon_d$:
\begin{equation}
\label{eqn80}
  \epsilon_{d} =  |\mu - \mu_d| \leq \beta/2 \left(\frac{\Delta x}{N_d}\right)^2 
\end{equation}

We now turn to the second error which is introduced by performing the classical Monte-Carlo on a digital computer, namely the sampling error. To ensure consistency with the quantum Monte-Carlo case, we consider the sampling technique described by Applegate and Kannan \cite{logconcave} according to which sampling from $p(x)$ requires 
\begin{equation}
\label{eqn81}
\mathcal{O}\left(\log \frac{1}{\epsilon_{s_{max}}}+ \log N_d \right)
\end{equation}
operations to achieve a maximum sampling error of $\epsilon_{s_{max}}$.\\
\indent Thus the discrepancy in the expectation of the samples, $\mu_s$, and the actual mean of the discretised distribution, $\mu_d$, can be bounded by considering the case in which each of the discretised probability masses errs by the maximum amount $\epsilon_{s_{max}}$, thus meaning that the entire distribution is inflated (or deflated) by a factor $1+ \epsilon_{s_{max}}$ (or $1- \epsilon_{s_{max}}$) and so the mean is increases (or decreased) accordingly. So we get:
\begin{equation}
\label{eqn190}
    (1-\epsilon_{s_{max}}) \mu_d < \mu_s < (1+ \epsilon_{s_{max}}) \mu_q
\end{equation}
Of course, this is a bounding case, as it will remain the case that the probabilities in the distribution with sampling errors must still sum to one, hence why they are strict inequalities in (\ref{eqn190}), as these bounds cannot actually be saturated.\\
\indent Returning to (\ref{eqn170}), we can now see that we are sampling from a distribution whose mean differs from the mean of the (continuous) distribution of interest by the discretisation and sampling errors ($\epsilon_d$ and $\epsilon_s$ respectively), and we define $\mu' = \mu + \epsilon_d + \epsilon_s$ such that it is the mean of the distribution we are sampling from. We can therefore re-express (\ref{eqn170}):
\begin{align}
\hat{\epsilon}^2  = & \mathbb{E}\left( \left( \mu' + \epsilon_d + \epsilon_s - \frac{1}{N_s} \sum_{j=1}^{N_s} X_j  \right)^2 \right) \nonumber \\
 = & \mathbb{E} \left( \left( \mu' - \frac{1}{N_s} \sum_{j=1}^{N_s} X_j \right)^2 \right) + (\epsilon_d + \epsilon_s)^2 \nonumber \\
& + 2(\epsilon_d + \epsilon_s) \mathbb{E}\left( \mu' - \frac{1}{N_s} \sum_{s=1}^{N_s} X_j  \right) \nonumber \\
 = & \mathbb{E} \left( \left( \mu' - \frac{1}{N_s} \sum_{j=1}^{N_s} X_j \right)^2 \right) + (\epsilon_d + \epsilon_s)^2 \nonumber \\ 
\label{eqn200}
 \leq & \mathbb{E} \left( \left( \mu' - \frac{1}{N_s} \sum_{j=1}^{N_s} X_j \right)^2 \right) + ( \epsilon_{d_{max}} + \epsilon_{s_{max}})^2
\end{align}
Thus we have expressed the RMSE as the actual error that is suppressed by the Monte-Carlo integration (i.e., the first term in the right-hand side (RHS) of (\ref{eqn200})), and also the errors pertaining to the discretisation and sampling (i.e., the second term in the RHS of (\ref{eqn200})).\\
\indent The first term is the error suppressed by classical Monte-Carlo, i.e., such that $\mathbb{E} \left( \left( \mu' - \frac{1}{N_s} \sum_{j=1}^{N_s} X_j \right)^2 \right) \in \Theta(\frac{1}{N_s})$; more pertinently, we can see from the second term in (\ref{eqn200}) that to achieve RMSE $\hat{\epsilon}$ it suffices to set up the digital computation such that $\epsilon_{s_{max}}, \epsilon_{d_{max}} \in \Theta(\hat{\epsilon})$. From (\ref{eqn80}) and  (\ref{eqn81}), we have that
\begin{equation}
    \label{eqn201}
    N_d \in \text{Poly} \left( \frac{1}{\hat{\epsilon}} \right)
\end{equation}
suffices to achieve this. Finally, we recall that $N_d$ is the number of intervals in the discretised distribution, and thus letting $n= \log_2 N$ we have that $\mathcal{O}(n)$ bits are required. All of the operations required in classical Monte-Carlo, such as summing up samples, are polynomial in the number of bits (noting that the complexity of generating a single sample, (\ref{eqn81}), incorporates operations on the number of bits required to digitally express a sample to the required accuracy) and so the computational overhead is 
\begin{equation}
    \label{eqn202}
    \text{Poly}(n) = \text{Poly} \log N = \text{Poly} \log \left( \frac{1}{\hat{\epsilon}} \right)
\end{equation}
Thus, when we include the sample complexity, we get an overall computational complexity
\begin{equation}
    \label{eqn203}
    \mathcal{O} \left( \left( \frac{1}{\hat{\epsilon}} \right)^2 \text{Poly} \log \left( \frac{1}{\hat{\epsilon}} \right) \right) = \tilde{\mathcal{O}} \left( \left(  \frac{1}{\hat{\epsilon}} \right)^2 \right)
\end{equation}

\end{document}